\documentstyle[proc]{rspublic}
\input epsf.sty
\newcommand{\ket}[1]{| #1 \rangle}
\begin{document}
\title[Gates and Circuits]{Quantum Gates and Circuits}
\author[D. P. DiVincenzo]{David P. DiVincenzo}
\affiliation{IBM Research Division, Thomas J. Watson Research Center,\\
Yorktown Heights, NY 10598 USA}
\maketitle
\begin{abstract}
A historical review is given of the emergence of the idea of the quantum
logic gate from the theory of reversible Boolean gates.  I highlight the
quantum XOR or controlled NOT as the fundamental two-bit gate for quantum
computation.  This gate plays a central role in networks for quantum
error correction.
\end{abstract}
\section{Introduction}

In this contribution I survey some topics of current interest in the
properties of quantum gates and their assembly into interesting
quantum circuits.  It may be noted that this paper, like a large
fraction of the others to be found in this volume of contributions to
the ITP conference on {\em Quantum Coherence and Decoherence}, is
about quantum computation and, apparently, not about quantum coherence
at all.  Let me assure the reader that it has not been the intention
of me or my coorganizer Wojciech Zurek to engage intentionally in false
advertising in the title of this conference.  In fact, it is my view,
and, I hope, a dominant view in this field, that quantum computation
has everything to do with quantum coherence and decoherence.  Indeed,
I think of quantum computation as a very ambitious program for the
exploitation of quantum coherence.  Many of us at this meeting believe
that quantum computers will be, first and foremost, the best tools
that we have ever invented for probing the fundamental properties of
quantum coherence.  It is thus that I justify the legitimacy of what
will be found in this volume, all under the heading of quantum
coherence.

In this paper I will first give a brief historical survey of the role
of reversibility in the theory of computation, and the early
discussion of gate and circuit constructions in reversible
computation.  I will review how these gates were re-interpreted as
quantum operations at the outset of discussions on quantum
computation.  I will spend a lot of time reviewing many properties of
what we now view as the most important gate for quantum computation,
the two-bit quantum XOR gate (or controlled NOT).  It has a role to
play in quantum measurement, in creation and manipulation of
entanglement, and in the currently popular schemes for quantum error
correction; it has also been the object of experimental efforts to
implement a two-bit quantum gate using precision spectroscopy.
Furthermore, it is the fundamental two-bit gate in the ``universal''
quantum gate constructions which we have introduced.

I will conclude with a couple of special topics involving the quantum
XOR, in particular, I will describe a procedure for turning the
group-theoretic description of orthogonal quantum codes into a gate
array, involving XORs and Hadamard-type one-bit gates, which decodes
(or encodes) from a coded qubit (or block of qubits) to the ``bare''
versions of these qubits, while correcting errors in them.

\section{Historical survey of elementary gates}

I choose to begin my history with the work of my esteemed colleague
Charles Bennett in the early '70's (Bennett 1973).  (The parallel work
of Lecerf was published earlier, but had no influence on the
development of the field.)  Inspired by the work of Landauer to ask
questions about the constraints which physics places on computation,
in 1973 Bennett announced that in one important respect physics does
{\em not} constrain computation: in particular, he found that
computation can be done {\em reversibly}.  At that time, workers were
focussed on one consequence of this discovery, namely that the
expenditure of free energy per step of boolean computation has no
lower limit.  This observation, while it has had no practical
implications in the intervening twenty-three years, clearly indicates
a profoundly important feature of the far future of computation.
\begin{figure}
\epsfxsize=13cm
\leavevmode
\epsfbox{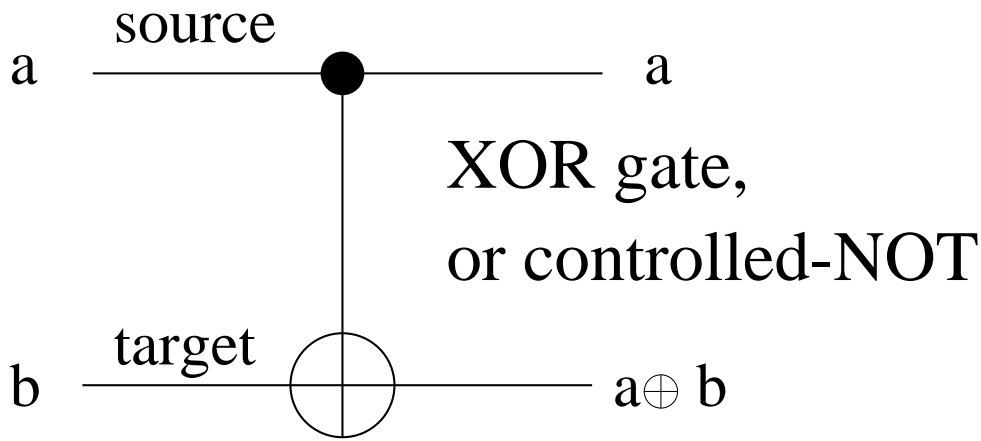}
\caption{The reversible XOR gate, or the controlled NOT.  I use the
notation first introduced by Toffoli (1980).  The top bit is
conventionally called the ``source,'' and the bottom one is called the
``target.''}\label{XORfig}
\end{figure}

For the present more modest tale, however, Bennett's discovery had a
number of more immediate intellectual consequences.  In the late 70's
Tom Toffoli (Toffoli 1980), inspired by the Bennett reversibility result,
investigated how reversible computing could be done in the traditional
language of Boolean logic gates.  He showed that a set of modified
gates could be used in place of the traditional Boolean logic gates
like AND, OR, etc.  One of these, which has turned out to be of
central importance in the subsequent quantum-gate work is the gate
shown in Fig.~\ref{XORfig}.  This gate, known in the literature as the
reversible XOR gate or the controlled-NOT gate, has the action
indicated: the $b$ bit is transformed to the exclusive-or (addition
modulo 2), while the $a$ bit is unchanged.  The simple retention of
the $a$ bit makes the gate reversible --- the input is a unique
function of the output.
\begin{figure}
\epsfxsize=13cm
\leavevmode
\epsfbox{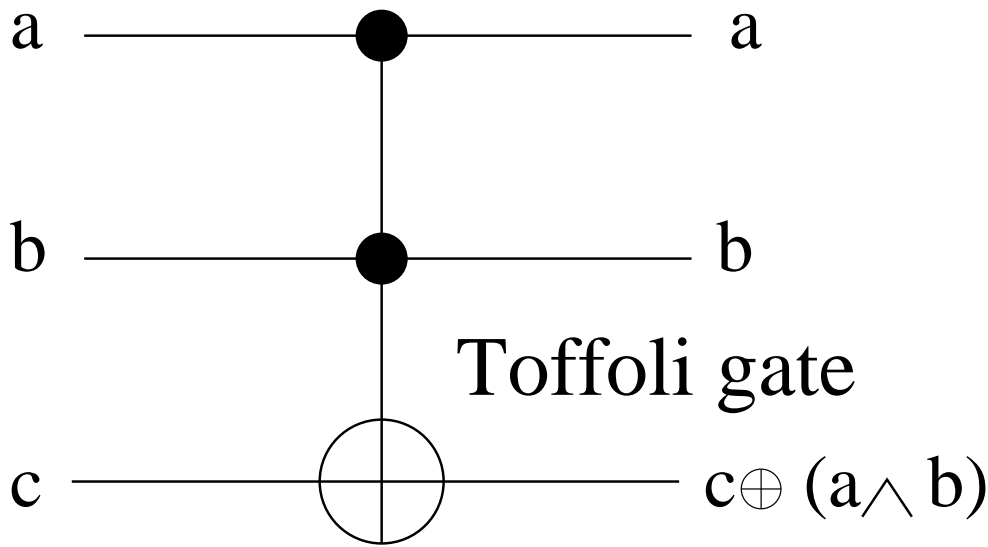}
\caption{The three-bit Toffoli gate (Toffoli 1980), shown to be
universal for reversible Boolean logic.  The action of the gate on
the three input bits is indicated.}\label{AND}
\end{figure}

The XOR gate is not ``universal'' for Boolean computation.  A
``universal'' logic gate is one from which one can assemble a circuit
which will evaluate any arbitrary Boolean function.  In ordinary
(irreversible) Boolean logic, NAND (or AND supplemented by NOT) is one
choice for the universal gate.  Toffoli sought another reversible gate
which could play the role of a universal gate for reversible circuits.
He found what we now call the ``Toffoli gate'', symbolized in
Fig.~\ref{AND}.  His gate requires three bits, and it is easy to show
(by exhaustive search, say) that any universal reversible boolean
logic gate must have at least three bits.  In essence, this gate is an
AND gate in which both input bits are saved; as Fig.~\ref{AND}
indicates, bits $a$ and $b$ are unchanged, while bit $c$ is
``toggled'' by $a\wedge b$.

It is easy to prove that the Toffoli gate is universal: Consider an
ordinary Boolean circuit using only NANDs.  Each of these may be
replaced by a Toffoli gate (setting $c=1$ at the input produces the
NAND function).  The only remaining difficulty is one of efficiency:
with this prescription, the number of extra bits introduced into the
circuit would grow linearly with the number of gates in the circuit
$T$, an undesirable efficiency penalty.  However, Bennett (1989)
introduced a ``pebbling'' technique in which the extra work bits are
reversibly erased and reused in stages throughout the operation of the
circuit; he showed that the number of bits can be arranged to increase
by just the factor $\epsilon 2^{1/\epsilon}\ln T$, at the cost of an
increase in the time of operation from $T$ to just order
$T^{1+\epsilon}$, for any $\epsilon>0$ (see also (Levine \& Sherman
1990)).  Thus the theory says that a Boolean circuit can always be
made reversible with little cost in efficiency.

The next milestone in this history of logic gates occurred a few years
later, when several workers (Benioff 1982; Feynman 1985) recognized
that the Hamiltonian time evolution of an isolated quantum system is a
reversible dynamics which may be made to mimic the steps of a
reversible Boolean computation.  I will summarize these ideas using
the language which we now use for these things.  This first step
towards quantum computation requires a reinterpretation at a very
basic level of the meaning of the reversible gates that I have
introduced above.  Let me speak first about the "quantization" of the
XOR gate above.  The two input bits are interpreted as the states of
two quantum two-level systems, with a correspondence made between the
0 state of the bit and an arbitrarily chosen basis state labeled
$|0\rangle$ of the quantum system, and between the 1 state and the
orthogonal state $|1\rangle$ of the two-level system.  The term {\em
qubit} has been coined to denote these two-level quantum-mechanical
states which play the role of bits.

With this available state space, the action of the quantum XOR is
simply described: it is a Hamiltonian process which maps the two-qubit
basis states according to the XOR truth table, viz.,
$|00\rangle\rightarrow |00\rangle$, $|01\rangle\rightarrow|01\rangle$,
$|10\rangle\rightarrow|11\rangle$, $|11\rangle\rightarrow|10\rangle$.

The statement that there exists a quantum Hamiltonian which accomplishes
the above mapping, presumably within some definite time $t$, of course
implies a great deal more about the time evolution of the quantum system;
these additional quantum properties were not considered by the original
authors, but were discussed first a few years later by Deutsch (1985).
Deutsch's observation is that the mappings on these basis states
uniquely specify the dynamics of an {\em arbitrary} initial quantum
state, simply on account of the linearity of the Schroedinger equation.
In this way of thinking, the time evolution of the quantum XOR may
be stated in a single line as
\begin{equation}
\alpha|00\rangle+\beta|01\rangle+\gamma|10\rangle+\delta|11\rangle
\rightarrow
\alpha|00\rangle+\beta|01\rangle+\delta|10\rangle+\gamma|11\rangle.
\label{XOR}
\end{equation}
This is to be true for arbitrary coefficients $\alpha$, $\beta$,
$\gamma$, and $\delta$ describing a normalized quantum state.

From this description it is easy to pass on to another one first
introduced in Deutsch (1989) which has been very common in discussions
of quantum gates, namely the description of the time evolution of
Eq. (\ref{XOR}) in terms of a unitary time-evolution matrix, which
relates the initial wavefunction coefficients to the final ones.  For
the quantum XOR the matrix is
\begin{equation}
U=\left(\begin{array}{llll}1&0&0&0\\0&1&0&0\\0&0&0&1\\0&0&1&0\end
{array}\right).\label{canonXOR}
\end{equation}

From this unitary operator it is possible to ``back out'' a
Hamiltonian which would implement the quantum gate, using the formula
\begin{equation}
U=\exp{i\int H(t)dt}.\label{ham}
\end{equation}
(Here I have omitted the time-ordered product for simplicity (Negele \&
Orland 1988).) It is possible to identify a time-independent
Hamiltonian, acting over a specific time interval, which will produce
the desired $U$; but in many spectroscopic applications it is actually
a time-dependent Hamiltonian which is used to accomplish the quantum
gate operation (Monroe {\em et al.} 1995).  In any case there is no
unique solution to Eq.~(\ref{ham}); there are many types of
Hamiltonians which may be used to implement this or any other quantum
gate.

I might observe on a personal note that it was when I first read
Deutsch's 1985 and 1989 papers (in 1994), and its discussion of the
quantum interpretation of reversible gates, that quantum computation
became compellingly interesting to me.

Before saying a few more words about the spectroscopic implementation
of the XOR, I would note that it is straightforward (Deutsch 1989;
DiVincenzo 1995a) to transcribe the Toffoli gate in the same way into
a unitary operator; in the eight-dimensional vector space spanned by
the basis states $|000\rangle$, $|001\rangle$, $|010\rangle$,
$|011\rangle$, $|100\rangle$, $|101\rangle$, $|110\rangle$, and
$|111\rangle$, the matrix is
\begin{equation}
U=\left(\begin{array}{llllllll}1&0&0&0&0&0&0&0\\0&1&0&0&0&0&0&0\\
0&0&1&0&0&0&0&0\\0&0&0&1&0&0&0&0\\0&0&0&0&1&0&0&0\\0&0&0&0&0&1&0&0\\
0&0&0&0&0&0&0&1\\0&0&0&0&0&0&1&0\end{array}\right).\label{Tmat}
\end{equation}

As I have noted in previous work (DiVincenzo 1995b) the quantum XOR
operation is embodied in some very old spectroscopic manipulations
which go under the general heading of double resonance operations.
In one of these, ``Electron-Nucleus Double Resonance'' (ENDOR), the
Hamiltonian of the relevant electron spin and the nuclear spin with
which it is interacting may be written as
\begin{equation}
H=g\mu_eH_0S_e^z+g\mu_NH_0S_N^z+JS_e^zS_N^z+H(t).
\end{equation}
\begin{figure}
\epsfxsize=13cm
\leavevmode
\epsfbox{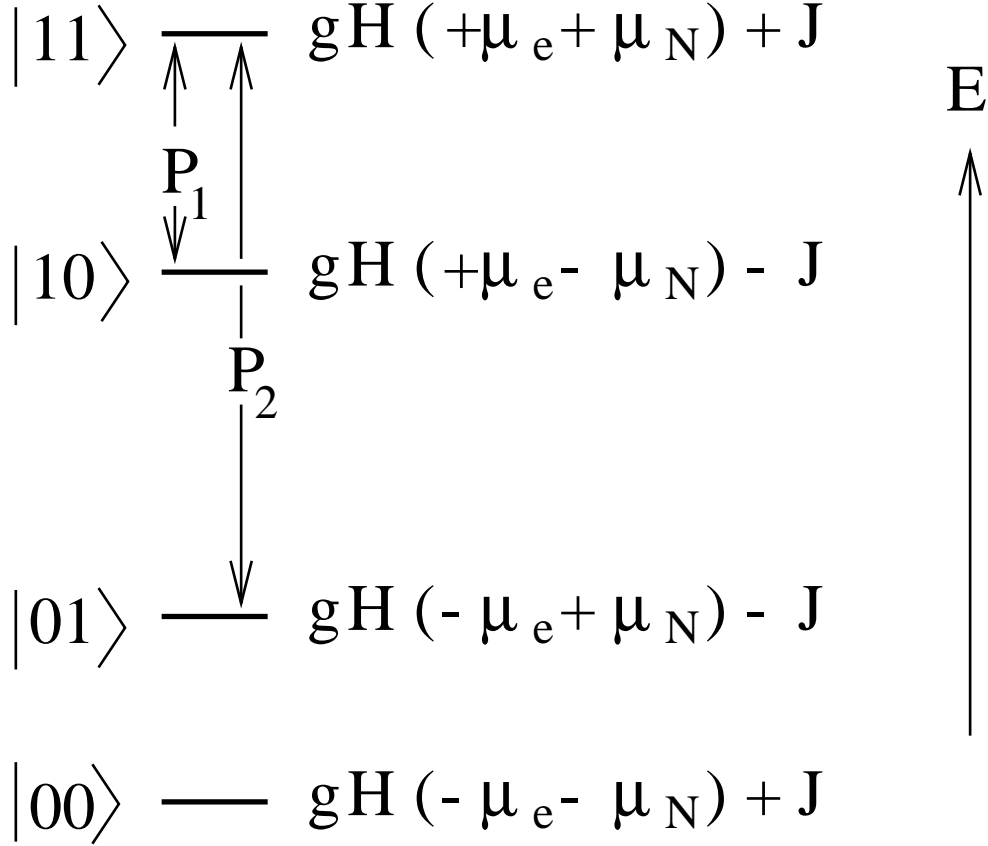}
\caption{Energy-level scheme and pair of inversions $P_1$ and $P_2$
used in the simplified version of Electron-Nucleus Double Resonance
(ENDOR), described as an implementation of XOR in the
text.}\label{spectra}
\end{figure}
In this slight simplification we take the interaction term between the
nucleus and the electron to have a simple Ising form (i.e., only
depending on the $z$ components of the spin operators).  The two
Zeeman interaction terms with DC magnetic field $H_0$ is standard.
The two-qubit gate is obtainable no matter what the form of the
interaction term, but it is easier to explain in this form.  The
time-dependent term will represent the spectroscopic protocol which
will actually accomplish the gate operation.  Before $H(t)$ is turned
on, the time-independent Hamiltonian has four distinct energy
eigenstates which are labeled by the four spin-up/spin-down
combinations of the spins (we assume a spin-1/2 nucleus); see
Fig.~\ref{spectra}.  $H(t)$, whose explicit form involves a pulse of
sinusoidally-varying magnetic field polarized in the X-Y plane (Baym
1969), is designed to have the effect of Rabi-flopping between
selected pairs of energy eigenstates in Fig.~\ref{spectra}; XOR is
accomplished by $180^\circ$ flops, or ``tips'' of the selected two
states.  In ENDOR there are two pulses involved, the first of which
pi-pulses the third and fourth levels, and the second of which
pi-pulses the second and fourth levels.  The unitary transformation
which this effects (in the standard basis) is
\begin{equation}
U=\left(\begin{array}{rrrr}1&0&0&0\\0&0&0&1\\0&-1&0&0\\0&0&-1&0
\end{array}\right).\label{badXOR}
\end{equation}
This differs from the XOR only in a couple of minor respects: First,
in addition to placing the result of the XOR in the first spin, it
also leaves the second spin in the initial state of the first spin.  A
simple modification of this pulse sequence can be made to avoid doing
this ``polarization transfer'', but it is amusing to note that this is
the very feature of ENDOR which has made it very valuable as a
spectroscopy --- there are many contexts in biology, chemistry, and
physics where it is useful to transfer the (high) polarization of a
electron to the (initially unpolarized) nucleus.  Actually, the XOR
action is an entirely ``unintended'' byproduct of performing ENDOR.
The second difference with the XOR, and a potentially serious one, is
that several of the phases in Eq.~(\ref{badXOR}) are different from
the canonical XOR of Eq.~(\ref{canonXOR}), in which all the phases are
zero (i.e., all the matrix elements are 0 or $+1$).  This does make
the ENDOR essentially different from XOR, as control of the quantum
phases is quite important in many quantum computation applications.
However, we have shown (Barenco {\em et al.} 1995) a variety of
methods by which gates with non-standard phases may be made to emulate
gates with standard phases, and such tricks are available in this case
as well.

Continuing the historical line one step further, I would like to
tabulate a number of properties of the XOR gate, which were mostly
first introduced in the Deutsch (1989) paper, that make the XOR of
central importance in many of the quantum computation constructions
which we consider today.

\begin{enumerate}
\begin{figure}
\epsfxsize=13cm
\leavevmode
\epsfbox{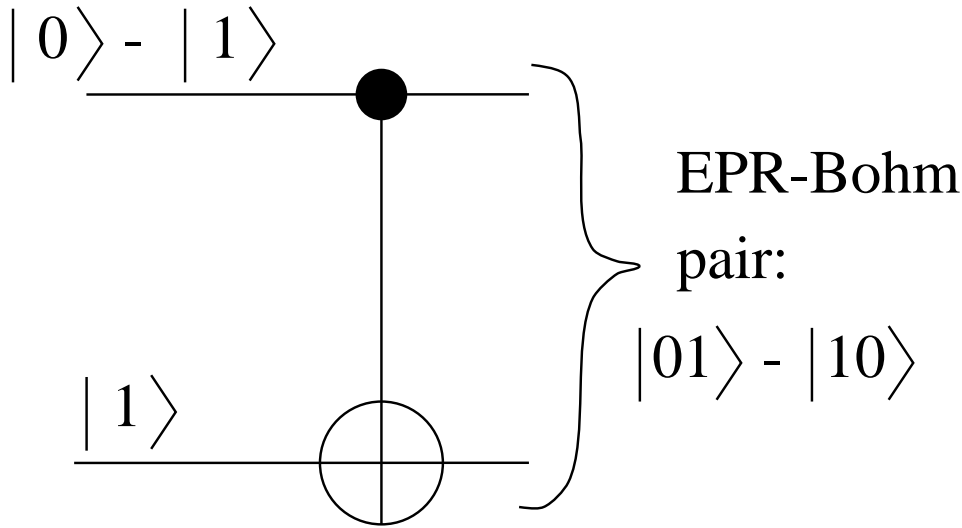}
\caption{XOR produces perfectly entangled quantum states from
unentangled ones.}\label{entang}
\end{figure}
\item The XOR is the idealized discrete operation for producing
entangled quantum states.  As Fig.~\ref{entang} indicates, a
particular product-state input to the gate as shown, using two states
from non-orthogonal bases (related by a Hadamard transform), produces
at the output the non-product state
$\frac{1}{2}(|01\rangle-|101\rangle)$, a state equivalent to an
Einstein-Podolsky-Rosen-Bohm pair.
\begin{figure}
\epsfxsize=13cm
\leavevmode
\epsfbox{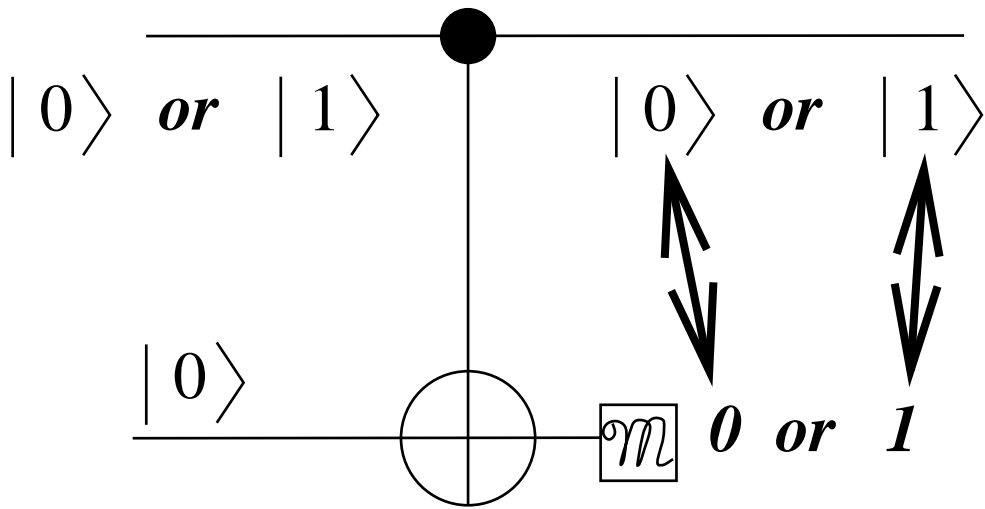}
\caption{XOR functions as an ideal non-demolition measurement apparatus
for a qubit.}\label{meas}
\end{figure}
\item As Deutsch (1989) termed it, the XOR also functions as a
``measurement gate.''  What he meant by this is illustrated in
Fig.~\ref{meas}: if the object is to measure the state of the upper
qubit (that is, whether it is in the $|0\rangle$ state or the
$|1\rangle$ state), we may XOR it with a second bit started in the
$|0\rangle$ state; then a measurement of the second bit will reveal
the desired outcome.  This may not appear to be much of an advantage
over measuring the first qubit directly.  However, it has the feature
of being a ``non-demolition'' measurement (Chuang \& Yamamoto 1996) in
which the original quantum state remains in existence after the
measurement.  Of course it only remains undisturbed if it started in
the $|0\rangle$ or the $|1\rangle$ state; if it started in a
superposition, then the state is ``collapsed'' by the measurement.
\begin{figure}
\epsfxsize=13cm
\leavevmode
\epsfbox{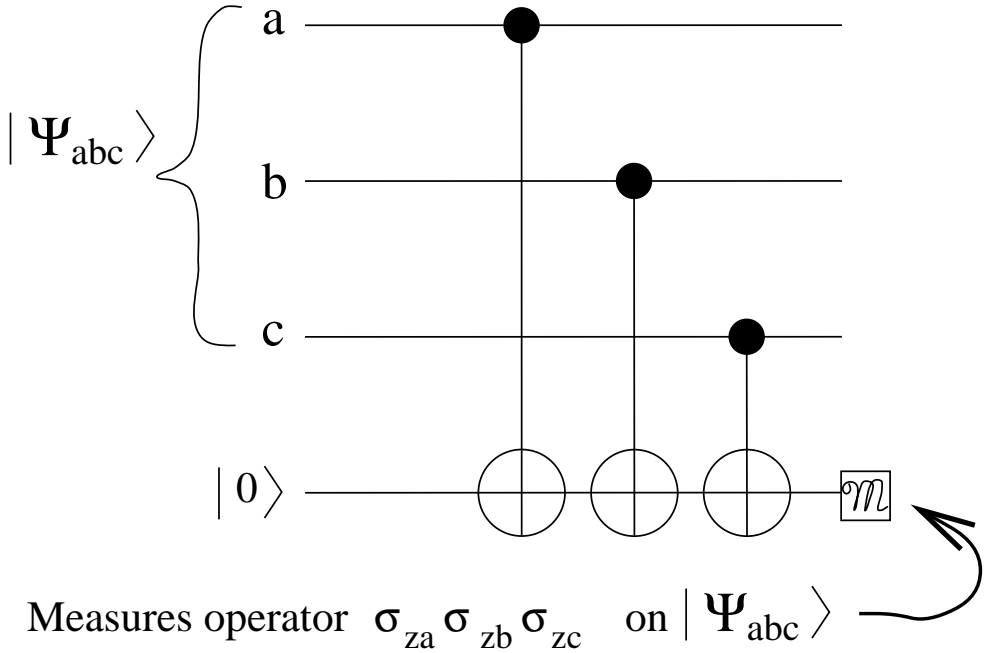}
\caption{A circuit of XORs can be used to do a non-demolition measurement
of the three-particle operator shown.}\label{QNDfig}
\end{figure}
\item To appreciate the real power of the non-demolition capability of
the XOR, consider the simple quantum circuit of Fig.~\ref{QNDfig}.
The effect of the three successive XORs followed by a measurement of
the target qubit is to accomplish a highly non-trivial non-demolition
measurement of the three-particle Hermitian operator $\sigma_{za}
\sigma_{zb}\sigma_{zc}$.  Previous discussions (Mermin 1990) of such
three-particle operators always assumed that it would necessarily be
done in a ``demolishing'' fashion where each of the one-particle
operators were measured separately.  As I will touch on a bit in
Sec. 4, this property forms the basis of the use of the XOR gate in
the implementation of error correction and hence in fault-tolerant
quantum computation. \item There are various applications in which we
use the simplicity of the XOR operation in several other bases.  If we
define (Steane 1996) a conjugate qubit basis by
$|\bar{0}\rangle=\frac{1}{\sqrt{2}} (|0\rangle+|1\rangle)$ and
$|\bar{1}\rangle=\frac{1}{\sqrt{2}} (|0\rangle-|1\rangle)$, then it is
easy to show that:
\begin{enumerate}
\item When both input qubits are considered in the conjugate basis,
then the effective gate action is an XOR but with the source and target
bits reversed.
\item If just the target bit is represented in the conjugate basis, then
the action of the XOR is completely symmetric on the two qubits, having
the form
\begin{equation}
U=\left(\begin{array}{rrrr}1&0&0&0\\0&1&0&0\\0&0&1&0\\0&0&0&-1\end{array}
\right).
\end{equation}
This ``phase-shift'' form of the gate is the one which has been discussed
in the cavity quantum-electrodynamic implementation of a two-bit quantum
gate (Turchette 1995).
\end{enumerate}
\end{enumerate}

\section{Another use of the XOR: universality for quantum gates}

I would like to feature separately another reason why we consider the
XOR gate so fundamental: we showed (Barenco {\em et al.} 1995) that the
XOR gate, when supplemented by a repertoire of one-bit quantum gates,
is sufficient to perform any arbitrary quantum computation.
Furthermore, as the constructions I am about to review show, many
important quantum computations are formulated quite naturally using
this repertoire.

In constructing our proof that this repertoire (XOR plus one-bit
gates) is ``universal'' for quantum computation in the sense that the
Toffoli gate above is universal for reversible Boolean computation, we
were able to make use of another important early discovery of
Deutsch (1989), which was that three-qubit quantum gates $U_D$ are
universal for quantum gate constructions, where $U_D$ has the
``double-controlled'' form
\begin{equation}
U_D=\left(\begin{array}{llllllll}1&0&0&0&0&0&0&0\\0&1&0&0&0&0&0&0\\
0&0&1&0&0&0&0&0\\0&0&0&1&0&0&0&0\\0&0&0&0&1&0&0&0\\0&0&0&0&0&1&0&0\\
0&0&0&0&0&0&u_{11}&u_{12}\\0&0&0&0&0&0&u_{21}&u_{22}\end{array}\right).
\label{D3}
\end{equation}
Here the $u_{ij}$'s constitute a generic $U(2)$ matrix.  Deutsch's
gate is a quantum generalization of the Toffoli gate, as the notation
of Fig.~\ref{AND} suggests.

\begin{figure}
\epsfxsize=13cm
\leavevmode
\epsfbox{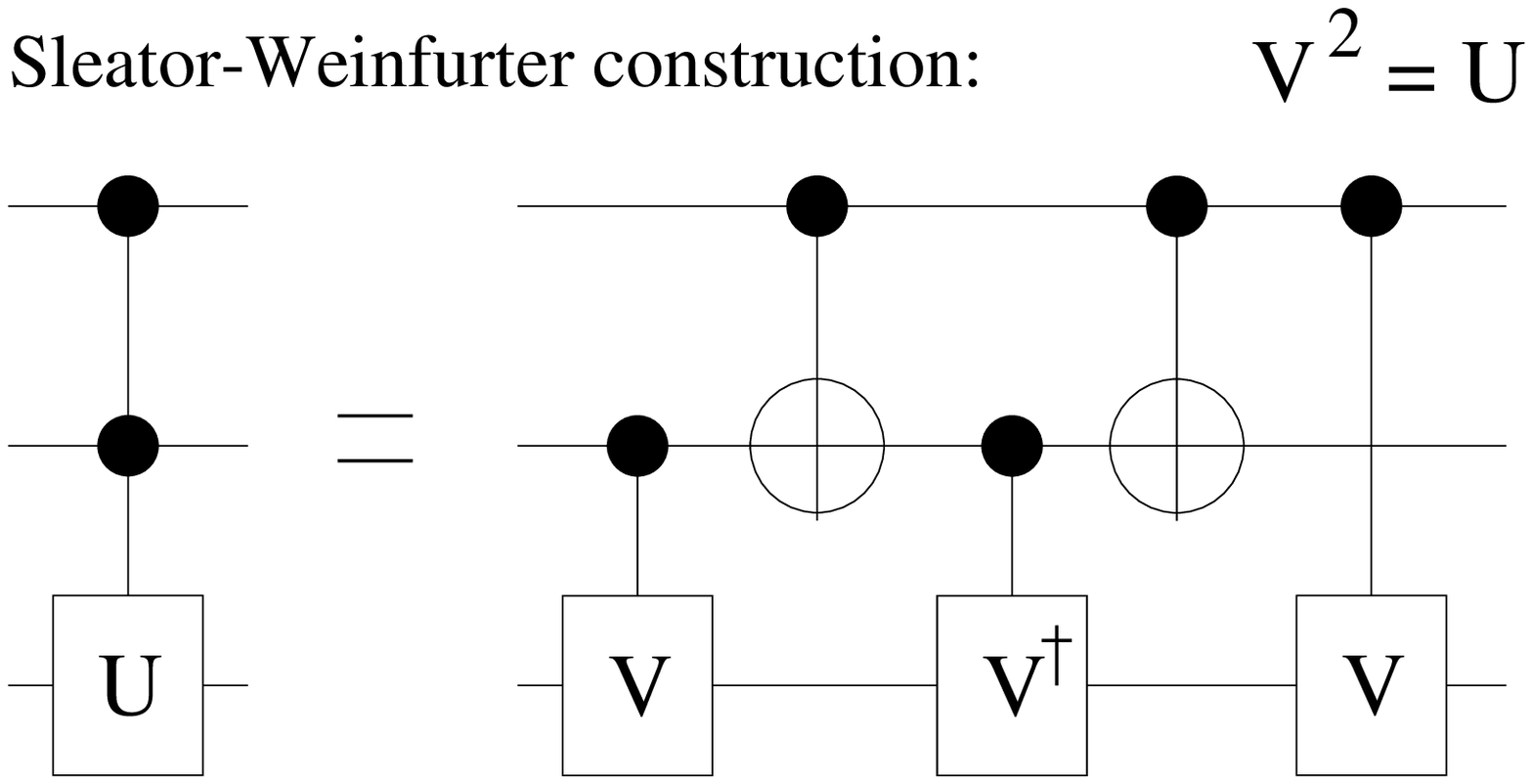}
\caption{The construction showing the Deutsch three-qubit gate of
Eq.~(3.1) can be broken down into a series of two-qubit gate
operations.}\label{S-W}
\end{figure}
It is fortunate for the prospects for the physical implementation of
quantum computation that, unlike in boolean reversible computation,
the Deutsch gate can indeed be broken down into simpler parts
(DiVincenzo 1995a; Deutsch {\em et al.} 1995; Lloyd 1995).  Probably the
simplest means of achieving this decomposition (Barenco {\em et al.}
1995) is shown in Fig.~\ref{S-W}.  The first step of the decomposition,
discovered by Sleator and Weinfurter (1995), utilizes two XOR
gates and three ``controlled-V'' gates, whose matrix description is
\begin{equation}
U_V=\left(\begin{array}{rrrr}1&0&0&0\\0&1&0&0\\
0&0&V_{11}&V_{12}\\0&0&V_{21}&V_{22}\end{array}\right).
\end{equation}
\begin{figure}
\epsfxsize=13cm
\leavevmode
\epsfbox{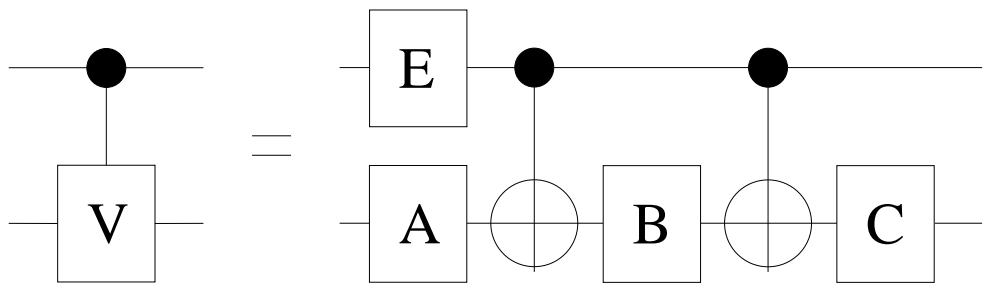}
\caption{Decomposition of the controlled-V gate into XORs and one-bit
gates.}\label{moreb}
\end{figure}
Here $V$ is a $U(2)$ matrix such that $V^2=U$.  We further showed how
these two-bit controlled-V gates could be further broken down, as
shown in Fig.~\ref{moreb}.  Here $A$, $B$, $C$, and $E$ are one bit
gates for which we (Barenco {\em et al.} 1995) have obtained explicit
formulas.

\begin{figure}
\epsfxsize=13cm
\leavevmode
\epsfbox{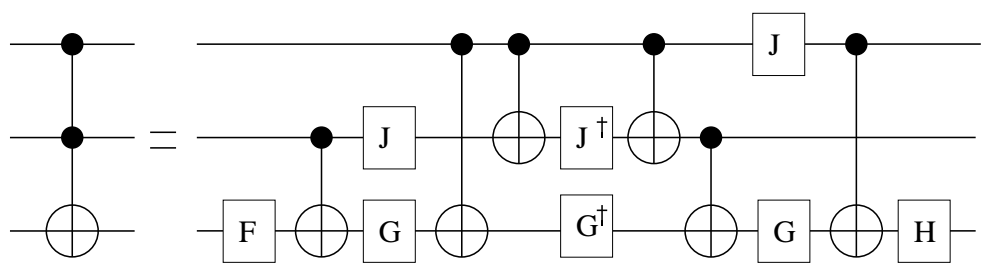}
\caption{Simplest known decomposition of the Toffoli gate into 6 XOR
gates and 8 one-bit gates (specified in Eq.~(3.3)).}\label{Tbreak}
\end{figure}
Thus, the circuit elements which would be needed in any quantum
computation in which we are currently interested can be readily be
simulated by short sequences of two-bit gates.  For instance, the
Toffoli gate, which would be the basis of much of the ordinary boolean
logic which is needed for large sections of, for example, Shor prime
factoring, can be obtained with just 6 XORs and 8 one-bit gates by
using the constructions above; this is shown in Fig.~\ref{Tbreak}.
The unitary operators for the one-bit gates in this construction are
\begin{equation}
\begin{array}{ll}F=\left(\begin{array}{rr}
e^{i\pi/4}\cos{\pi/8}&e^{i\pi/4}\sin{\pi/8}\\
-e^{-i\pi/4}\sin{\pi/8}&e^{-i\pi/4}\cos{\pi/8}\end{array}\right)&
G=\left(\begin{array}{rr}
\cos{\pi/8}&-\sin{\pi/8}\\\sin{\pi/8}&\cos{\pi/8}\end{array}\right)\\
H=\left(\begin{array}{cc}e^{-i\pi/4}&0\\0&e^{i\pi/4}\end{array}\right)&
J=\left(\begin{array}{cc}1&0\\0&e^{-i\pi/4}\end{array}\right)
\end{array}
\end{equation}
It may be noted that in this construction the gates can be grouped
into a sequence of just five two-bit operations: first a 2-3
operation, then 1-3, 1-2, 2-3, and finally 1-3 (numbering the qubits
1-2-3 from the top).  Numerical simulations (DiVincenzo \& Smolin
1994; Smolin \& DiVincenzo 1996) have indicated that the Toffoli gate
can be obtained with no fewer than five two-bit quantum gates of any
type.

\begin{figure}
\epsfxsize=13cm
\leavevmode
\epsfbox{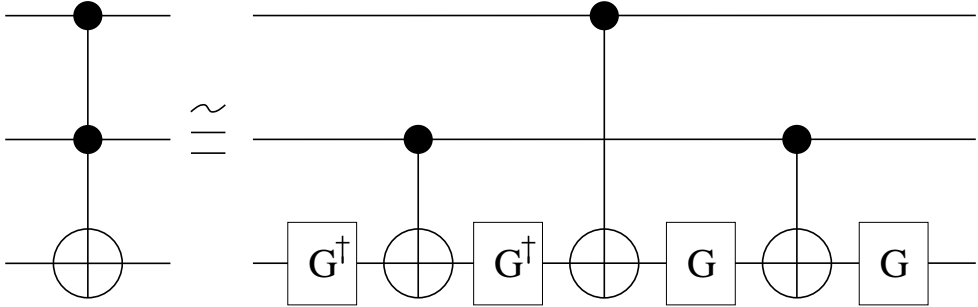}
\caption{Margolus's simplified Toffoli gate construction, if just
one of the quantum phases is allowed to be changed.}\label{almost}
\end{figure}
In a related result, Margolus has found (Barenco {\em et al.} 1995) an
``almost'' Toffoli gate which requires even less resources, as shown
in Fig.~\ref{almost}: just three XORs and four one-bit gates (three
two-bit gates overall). It is ``almost'' in the sense that one of the
matrix elements of the Toffoli gate in Eq. (\ref{Tmat}) is changed from
1 to $-1$ (the one corresponding to the $|100\rangle$ state).  This is
not generally acceptable for quantum computation, where all the phases
must be correct; however, we have noted (Barenco {\em et al.} 1995;
Cleve \& DiVincenzo 1996) that in many quantum programs the Toffoli
gates appear in pairs, so that the ``wrong'' phase of the Margolus
construction can be arranged to cancel out.

\section{Gate constructions for quantum error correction}

I wish to briefly touch upon a few points about the recent developments
in error correction in quantum computation.  Other authors in this book,
and in many other papers, have given a very complete and rigorous
discussion of this topic.  Here I have only modest goals in mind:
for the most part, I want to point out a few ways in which error
correction uses and illuminates the ideas of quantum gate construction
which I have been discussing above.

\begin{figure}
\epsfxsize=13cm
\leavevmode
\epsfbox{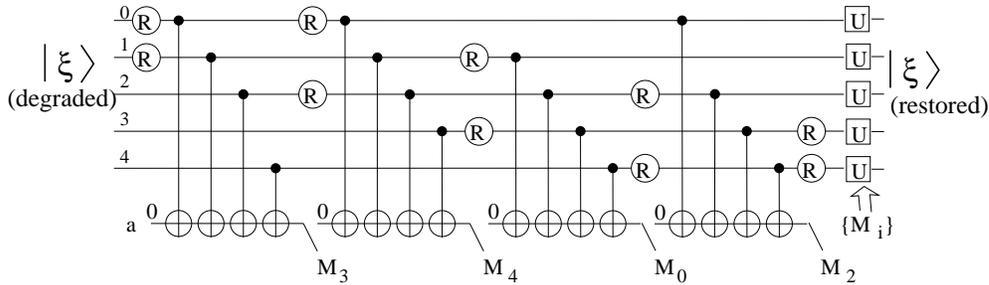}
\caption{Restoration network for the five-bit quantum error correcting
code (see DiVincenzo \& Shor (1996)).}\label{5bit}
\end{figure}
I begin immediately with a gate array shown in Fig.~\ref{5bit} which
is of considerable interest to me (DiVincenzo \& Shor 1996).  This is
one of the simplest ``restoration'' networks which works on a qubit
state
\begin{equation}
|\xi\rangle=a|0\rangle_L+b|1\rangle_L,\label{codeq}
\end{equation}
where
\begin{eqnarray}
\ket{0}_L &=& \ket{00000}\label{sym0}\\
&+& \ket{11000} +\ket{01100} +\ket{00110} +\ket{00011} +\ket{10001}
\nonumber \\
&-& \ket{10100} -\ket{01010} -\ket{00101} -\ket{10010} -\ket{01001}
\nonumber \\
&-& \ket{11110} -\ket{01111} -\ket{10111} -\ket{11011} -\ket{11101}
\nonumber
\end{eqnarray}
and
\begin{eqnarray}
\ket{1}_L &=& \ket{11111}\label{sym1}\\
&+& \ket{00111} +\ket{10011} +\ket{11001} +\ket{11100} +\ket{01110}
\nonumber \\
&-& \ket{01011} -\ket{10101} -\ket{11010} -\ket{01101} -\ket{10110}
\nonumber \\
&-& \ket{00001} -\ket{10000} -\ket{01000} -\ket{00100} -\ket{00010}
\nonumber .
\end{eqnarray}
Thus, $|\xi\rangle$ is coded into this two-dimensional subspace of the
five qubits entering the network from the left.  The network operates
in such a way that if $|\xi\rangle$ has been degraded by noise, but
only in such a way that no more than one of the five qubits is
significantly damaged (a condition which is true at short times for
most models of decoherence), then the network can restore the coded
qubit exactly to its pristine state.  We learned to our surprise from
the work of Shor (Shor 1995) that this would be true independent of
the state of the qubit in Eq.~(\ref{codeq}), that is, independent of
the coefficients $a$ and $b$; thus we seemed to arrive at some sort of
``analog'' error correction, which had been believed to be impossible.
Nevertheless, there is some deeper sense in which the error correction
scheme which this network embodies is really more akin to that
performed in digital computation.  I will leave this thorny issue for
other thinkers to try to articulate, but I hope to illustrate this by
explaining how this error correction network manages to operate.  At
the level of the recipe needed to obtain this and similar networks,
the procedure is very ``digital.''

Now, to return to the main point, the essence of the error restoration
of Fig.~\ref{5bit} is that it is performing four non-demolition
measurements, the outcomes of which are labeled $M_3$, $M_0$, $M_1$,
and $M_2$ in the figure.  If the one bit (Hadamard) rotations were
absent, then the effect of each measurement is easily understood in
classical Boolean logic (see Fig.~6).  Then the effect of the XORs,
say in $M_3$, is to collect up the parity of the bits 0, 1, 2, and 4,
saving this parity in the ancilla bit $a$.  Quantum mechanically,
the parity of these qubits is given by the eigenvalue of the Hermitian
operator $\sigma_{z0} \sigma_{z1}\sigma_{z2}\sigma_{z4}$, where the
$+1$ eigenvalue corresponds to even parity and the $-1$ eigenvalue to
odd parity.  Indeed, the effect of the four XOR gates is exactly to
perform the non-demolition measurement of this operator (see above for
a discussion of simple non-demolition measurement by one XOR gate).
The two $R$ gates preceding it can be thought of as basis changes for
qubits 0 and 1.  As it happens, this basis change is one which
interchanges $x$ and $z$ labels, i.e., $R\sigma_zR^\dagger=\sigma_x$,
$R\sigma_xR^\dagger=\sigma_z$, and $R\sigma_yR^\dagger=-\sigma_y$.
(Any sign changes in this basis transformation are irrelevant, since
these operators are just specifying the basis of a measurement.)
Thus, $M_3$ in Fig.~\ref{5bit} is actually a measurement of the
operator $\sigma_{x0} \sigma_{x1}\sigma_{z2}\sigma_{z4}$ on the state
$|\xi\rangle$.  The full set of four commuting operators which are
measured (in non-demolition fashion) by this network are
\begin{equation}
\begin{array}{rr}
M_3&\sigma_{x0}\sigma_{x1}\sigma_{z2}\sigma_{z4}\\
M_4&\sigma_{x1}\sigma_{x2}\sigma_{z3}\sigma_{z0}\\
M_0&\sigma_{x2}\sigma_{x3}\sigma_{z4}\sigma_{z1}\\
M_1&\sigma_{x3}\sigma_{x4}\sigma_{z0}\sigma_{z2}\end{array}.\label{QND}
\end{equation}
As has been shown in the beautiful work on the theory of
stabilizer-group codes by Gottesman (1996) and Calderbank {\em et al.}
(1996), the outcomes of these four measurements can completely
diagnose the error which has occurred on the coded quantum state
(assuming that the error affects no more than one qubit), so that a
final rotation (the $U$ gates of Fig.~\ref{5bit}) of the qubits which
have been determined to be in error will completely restore the coded
state to its pristine form before decoherence.

Almost all the quantum codes which have been discovered up until now
(the only exceptions that I know of are the single examples provided
by Leung {\em et al.} (1997) and Rains {\em et al.} (1997)) are
specified by generators like in Eq.~(\ref{QND}), which are given as
products of Pauli matrix operators.  It is clear that the method
exemplified in Fig.~\ref{QNDfig} for doing non-demolition measurements
can be easily generalized to any product of Pauli-matrix operators.
The rules can be summarized as follows: 1) make a basis change on each
of the qubits so that the Pauli operator is changed to a $\sigma_z$.
There are just two cases: If the operator is $\sigma_x$, then do the
basis change with $R$ as above.  If the operator is $\sigma_y$, then
change basis with
\begin{equation}
R'=\frac{1}{\sqrt{2}}\left(\begin{array}{rr}1&i\\i&1\end{array}\right)
\end{equation}
Which produces the transformation $R'\sigma_xR'^\dagger=\sigma_x$,
$R'\sigma_yR'^\dagger=-\sigma_z$, $R'\sigma_zR'^\dagger=\sigma_y$,
interchanging $y$ and $z$ as desired.  2) XOR each of the bits
involved into an ancilla bit.  Then a measurement of the ancilla bit
is the desired non-demolition measurement.  3) In most cases it is
desirable to undo the one-bit operations in order to restore the coded
qubits to their original basis.

The above discussion of the ``restoration'' network is largely of the
work reported in (DiVincenzo \& Shor 1996).  What has not been
previously discussed is how a very similar procedure leads to a
``decoding'' network for the same quantum code.  Decoding is like
restoration in that non-demolition measurements are performed to
determine the error syndrome.  Unlike in restoration, the mapping
takes the state from the coded form
$|\xi\rangle=a|0\rangle_L+b|1\rangle_L$ (or a noisy version thereof)
to the ``bare'' form $|\xi\rangle=a|0\rangle_L+b|1\rangle_L$.  Such
decoding has been discussed by Cleve \& Gottesman (1996), but the
procedure which I will now describes seems to be a little more compact
than theirs.

\begin{figure}
\epsfxsize=13cm
\leavevmode
\epsfbox{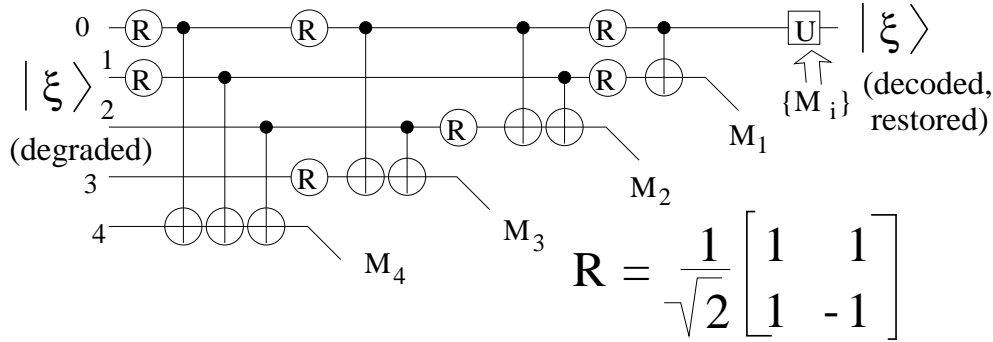}
\caption{A decoding network for the five-bit code, derivable
directly from the generators Eq.~(4.2).}\label{5decode}
\end{figure}
I will describe ``decoding'' by going through it for the same five bit
code as above.  The result is summarized in Fig.~\ref{5decode}.  I
will proceed by keeping track of the evolution of the four operators
describing the non-demolition measurement, Eq. (\ref{QND}), through
the basis changes made in the course of the decoding.  The first basis
change has exactly the same purpose as in the restoration network, to
bring the first operator into a form involving only $\sigma_z$'s:
\begin{equation}
\begin{array}{r}
\sigma_{z0}\sigma_{z1}\sigma_{z2}\sigma_{z4}\\
\sigma_{z1}\sigma_{x2}\sigma_{z3}\sigma_{x0}\\
\sigma_{x2}\sigma_{x3}\sigma_{z4}\sigma_{x1}\\
\sigma_{x3}\sigma_{x4}\sigma_{x0}\sigma_{z2}\end{array}\label{QND2}
\end{equation}
Now the next three XOR gates just accomplish a series of two-bit basis
changes.  The effect of these basis changes is easy to tabulate
(Bennett {\em et al.} 1996; Calderbank {\em et al.} 1996):
\begin{equation}
\begin{array}{rr}
U_{XOR}\sigma_{0s}\sigma_{0t}U_{XOR}^\dagger=\sigma_{0s}\sigma_{0t}&
U_{XOR}\sigma_{0s}\sigma_{xt}U_{XOR}^\dagger=\sigma_{0s}\sigma_{xt}\\
U_{XOR}\sigma_{0s}\sigma_{yt}U_{XOR}^\dagger=\sigma_{zs}\sigma_{yt}&
U_{XOR}\sigma_{0s}\sigma_{zt}U_{XOR}^\dagger=\sigma_{zs}\sigma_{zt}\\
U_{XOR}\sigma_{xs}\sigma_{0t}U_{XOR}^\dagger=\sigma_{xs}\sigma_{xt}&
U_{XOR}\sigma_{xs}\sigma_{xt}U_{XOR}^\dagger=\sigma_{xs}\sigma_{0t}\\
U_{XOR}\sigma_{xs}\sigma_{yt}U_{XOR}^\dagger=\sigma_{ys}\sigma_{zt}&
U_{XOR}\sigma_{xs}\sigma_{zt}U_{XOR}^\dagger=\sigma_{ys}\sigma_{yt}\\
U_{XOR}\sigma_{ys}\sigma_{0t}U_{XOR}^\dagger=\sigma_{ys}\sigma_{xt}&
U_{XOR}\sigma_{ys}\sigma_{xt}U_{XOR}^\dagger=\sigma_{ys}\sigma_{0t}\\
U_{XOR}\sigma_{ys}\sigma_{yt}U_{XOR}^\dagger=\sigma_{xs}\sigma_{zt}&
U_{XOR}\sigma_{ys}\sigma_{zt}U_{XOR}^\dagger=\sigma_{xs}\sigma_{yt}\\
U_{XOR}\sigma_{zs}\sigma_{0t}U_{XOR}^\dagger=\sigma_{zs}\sigma_{0t}&
U_{XOR}\sigma_{zs}\sigma_{xt}U_{XOR}^\dagger=\sigma_{zs}\sigma_{xt}\\
U_{XOR}\sigma_{zs}\sigma_{yt}U_{XOR}^\dagger=\sigma_{0s}\sigma_{yt}&
U_{XOR}\sigma_{zs}\sigma_{zt}U_{XOR}^\dagger=\sigma_{0s}\sigma_{zt}
\end{array}
\end{equation}
These equations are true modulo $\pm1,~\pm i$, which, as explained
above, are irrelevant for the error restoration or decoding.
$\sigma_0$ denotes the identity operator, and the subscripts $s$ and
$t$ denote operators on the source and target bits.  Using this table,
the effect of the first three XOR gates in Fig.~\ref{5decode} may be
easily worked out.  The four operators become:
\begin{equation}
\begin{array}{l}
\sigma_{z4}\\
\sigma_{x0}\sigma_{z1}\sigma_{x2}\sigma_{z3}\\
\sigma_{z0}\sigma_{y1}\sigma_{y2}\sigma_{x3}\sigma_{z4}\\
\sigma_{x0}\sigma_{z2}\sigma_{x3}\end{array}\label{QND2a}
\end{equation}
Note that the first operator can be measured by just making a conventional
measurement on spin 4; and that is what is done.  Of the other three
operators, only the third operator also involves spin 4.  Since all these
operators commute, the dependence can only be $\sigma_{z4}$
($\sigma_{x4}$ and $\sigma_{y4}$ would anticommute).  It is always
possible to create new measurement operators by mutiplying two
operators together (Calderbank {\em et al.} 1996), and if we do
so here, the dependence on spin 4 is eliminated in the third operator.
Because of the commuting condition, the spin-4 dependence can always
be eliminated from all the measurement operators except one.  Thus,
after measurement of qubit 4, qubits 0-3 still remain, and it still
remains to measure the operators
\begin{equation}
\begin{array}{l}
\sigma_{x0}\sigma_{z1}\sigma_{x2}\sigma_{z3}\\
\sigma_{z0}\sigma_{y1}\sigma_{y2}\sigma_{x3}\\
\sigma_{x0}\sigma_{z2}\sigma_{x3}\end{array}\label{QND2b}
\end{equation}
These can be dealt with in turn in exactly the same way.  Choosing
to simplify the last of these operators, the next set of one $R$
operation and two XORs lead to new operators
\begin{equation}
\begin{array}{l}
\sigma_{z0}\sigma_{z1}\sigma_{x2}\\
\sigma_{y0}\sigma_{y1}\sigma_{x2}\sigma_{z3}\\
\sigma_{z3}\end{array}\label{QND3}
\end{equation}
Now the second measurement becomes a measurement of the last remaining
operator.  Repeating this, the operators remaining to be measured are
\begin{equation}
\begin{array}{l}
\sigma_{z0}\sigma_{z1}\sigma_{x2}\\
\sigma_{y0}\sigma_{y1}\sigma_{x2}\end{array}\label{QND4}
\end{equation}
Now we choose to simplify the first remaining operator with the next set
of one $R$ operation and two XORs leading to the modified operators
\begin{equation}
\begin{array}{l}
\sigma_{z2}\\
\sigma_{x0}\sigma_{x1}\sigma_{z2}\end{array}\label{QND5}
\end{equation}
Measuring qubit two gives the next bit of the syndrome.  We finally have
just one remaining operator to measure,
\begin{equation}
\sigma_{x0}\sigma_{x1}\label{QND6}
\end{equation}
which is accomplished by the final two $R$s and single XOR in the
circuit.  Finally, a one-bit rotation conditional on the measured
syndrome bits is applied, just as in the restoration network of
Fig.~\ref{5bit} (DiVincenzo \& Shor 1996), to obtain the error
corrected, ``bare'' qubit.

This procedure provides a much more systematic procedure for obtaining
the restoration network than in the original work (Laflamme {\em et
al.} 1996; Bennett {\em et al.} 1996), although it is not clear that
the optimal network could be obtained by this procedure (Braunstein
\& Smolin 1997).  The procedure given above requires $O(n^2)$ gates,
as in the decoding networks proposed by Cleve \& Gottesman (1996);
however, the present procedure is superior in that the decoding and
the collection of the error syndrome are done simultaneously.  It
should also be noted that Fig.~\ref{5decode} can be converted directly
into an encoding network by inversion, with the final $U$ operation
removed and the measurements replaced by prepared $|0\rangle$ states.

It is also worthwhile to recall that this network, applied bilaterally
to both halves of five corrupted EPR pairs, results in the purification
of these pairs (Bennett {\em et al.} 1996).  The stabilizer-group theory
which we use now shows us that it is unnecessary ever to use unilateral
gate operations in the purification of Bell states, which was not clear
in our original work.  We see that each stage of non-demolition measurement
illustrated above is equivalent to one step of what we referred to as
``one-way hashing'' in our original paper (Bennett {\em et al.} 1996).
In our original language, each sequence of $R$s and XORs accumulates
a generalized parity bit of a subset of the amplitude and phase bits
specifying the Bell states.  For completeness, I note here, in the
notation of the original paper, the ``random'' parity strings
implemented in the network of Fig.~\ref{5decode}:
\begin{equation}
\begin{array}{l}
s_1=10 10 01 00 01\\
s_2=10 00 01 10\\
s_3=01 01 10\\
s_4=10 10.\end{array}
\end{equation}

Finally, I would like to say a few words about the relation between the
non-demolition measurement operators appearing in the stabilizer code
theory and the operators considered some seven years ago by Mermin
(Mermin 1990) in his study of states which violate the
Einstein-Podolsky-Rosen (EPR) notion of locality without the requirement
of Bell inequalities.  I leave most of this discussion to my paper
with Asher Peres (DiVincenzo \& Peres 1997) on this subject.

Mermin gave the simplest form of the states introduced by Greenberger,
Horne and Zeilinger (see Zeilinger {\em et al.} 1997); one of the states
he considered was
\begin{equation}
\Psi_{GHZ}=\frac{1}{\sqrt{2}}(|000\rangle-|111\rangle)
\end{equation}
This is obviously a highly entangled state; it is also a kind of ``code''
state, in that you might view the 000 and 111 as being simple triple
repetition codes standing for 0 and 1.  The EPR paradox which was brought
out by Mermin involve the following four operators:
\begin{equation}
\begin{array}{lr}
\sigma_{x1}\sigma_{y2}\sigma_{y3}&(=1)\\
\sigma_{y1}\sigma_{x2}\sigma_{y3}&(=1)\\
\sigma_{y1}\sigma_{y2}\sigma_{x3}&(=1)\\
\sigma_{x1}\sigma_{x2}\sigma_{x3}&(=-1)\end{array}\label{GHZ}
\end{equation}
The similarity with the measurement operators should already be evident.
These are again a set of commuting operators involving the Pauli
operators on the set of particles.  Unlike in error correction, these
operators are not all independent; the fourth one is the negative of the
product of the first three.  Of course, in the error correction scheme
products of the measurement operators also produce valid syndrome
measurements.  The appearance of a redundant operator is actually
quite crucial to the contradiction which Mermin draws, as I will
explain in a moment.  As in error correction, the four error operators
are eigenoperators on the state $\Psi_{GHZ}$; so indeed $\Psi_{GHZ}$
is exactly like the protected subspace in the error correction scheme.

The current development of error correction has diverged from Mermin's
work in what one does with the set of operators.  Rather than doing a
non-demolition measurement of the whole set of operators, Mermin
envisions a thought experiment in which a ``demolishing'' experiment
is done of one of the four operators on each run of an experiment
which begins with a perfect GHZ state.  (No noise is considered in
the Mermin setup.)  In this ``demolishing'' experiment the three
particles are imagined to be widely separated, and the operators are
measured by measuring each of the one-particle measurements separately;
this is followed by a classical comparison of the three measured results.
GHZ were constructing cases in which one could contradict the assertion
of ``hidden variable'' theories of quantum mechanics (the viewpoint of
EPR) that the values of operators on one particle which can be determined
by examination (i.e., measurement) of other, separated particles must
constitute ``elements of reality'', and thus have values which were
preset at the time that the quantum state was prepared.

Mermin pointed out that this assertion is contradicted in his GHZ
thought experiment.  First, he notes that every Pauli operator on
every particle conforms to the definition of an element of reality,
and thus should have a preset value for the GHZ state.  Then if he
multiplies the preset values of the four operators in Eq.~(\ref{GHZ})
the value must always be $+1$, since every operator appears exactly
twice and each can only be preset to values $\pm 1$.  However,
obviously the product of the actual four measurements from the
discussion above is $-1$.  This is the flat contradiction of hidden
variables which Mermin pointed out.

We see from this example that much of the machinery of quantum error
correction was actually latent in this previous work on the
foundations of quantum mechanics; we (DiVincenzo \& Peres 1997) have
found that all the presently-proposed quantum error correcting codes
provide exactly the same type of contradiction of hidden-variable
theory as Mermin found for the GHZ state.  This should not surprise
us, given the close relation, as exemplified in much of the work at
this conference, between quantum computation and the foundations of
quantum theory.  I expect that this association will continue to
produce fruitful results in the future.

\begin{acknowledgments}

I thank the other authors who collaborated with me on the work
described here: A. Barenco, C. Bennett, R. Cleve, N. Margolus,
A. Peres, P. Shor, T. Sleator, J. Smolin, H. Weinfurter and
W. Wootters.

\end{acknowledgments}

\label{lastpage}
\received{Received <date of receipt>;
          revised <date of revision>;
          accepted <date of acceptance>}
\end{document}